\documentclass[aps,twocolumn,prb,showpacs,amsmath,amssymb]{revtex4}
\usepackage{graphicx}
\usepackage{dcolumn}
\usepackage{bm}
\usepackage{amstext}
\usepackage{bbding}
\usepackage{pifont}
\usepackage{natbib}

\begin{document}

\preprint{AIP/123-QED}

\title{Optimal configuration of microstructure in ferroelectric materials
by  stochastic optimization}

\author{K.P. Jayachandran}
 \email{jaya@dem.ist.utl.pt}
\author{J.M. Guedes}
\author{H.C. Rodrigues}
\affiliation{IDMEC, Instituto Superior T{\'e}cnico , Technical University of Lisbon, Av. Rovisco Pais, 1049-001 Lisbon , Portugal}


\begin{abstract}
An optimization procedure determining the ideal configuration at the microstructural
level of ferroelectric (FE) materials is applied to maximize piezoelectricity. Piezoelectricity in ceramic FEs differ significantly from that of single crystals because of the presence of crystallites (grains)  possessing crystallographic axes aligned imperfectly. The piezoelectric properties of a polycrystalline (ceramic) FE is inextricably related to the grain orientation distribution (texture). The set of combination of variables, known as solution space, which dictates the texture of a ceramic is unlimited and hence the choice of the optimal solution which maximizes the piezoelectricity is complicated. Thus a stochastic global optimization combined with homogenization is employed for the identification of the optimal granular configuration of the FE ceramic microstructure with optimum piezoelectric properties. The macroscopic equilibrium piezoelectric properties of polycrystalline FE is calculated using mathematical homogenization at each iteration step. The configuration of grains characterised by its orientations at each iteration is generated using a randomly selected set of orientation distribution parameters. The optimization procedure applied to the single crystalline phase compares
well with the experimental data. Apparent enhancement of piezoelectric coefficient $d_{33}$ is observed in an optimally oriented BaTiO$_3$ single crystal. Based
on the good agreement of results with the published data in single crystals,
we proceed to apply the methodology in polycrystals. A configuration of  crystallites, simultaneously constraining the orientation distribution of the c-axis (polar axis) while incorporating ab-plane  randomness, which would multiply the overall piezoelectricity in ceramic BaTiO$_{3}$ is also identified. The orientation distribution of the c-axes is found to be a narrow Gaussian distribution centred around ${45^\circ}$. The piezoelectric coefficient in such a ceramic is found to be nearly three times as that of the single crystal.  Our optimization model provide designs for materials with enhanced piezoelectric performance, which would stimulate further studies involving materials possessing higher spontaneous polarization. 
\end{abstract}

\pacs{77.65.Bn,77.65.-j,77.80.-e,02.60.Pn}                             

\maketitle

\section{\label{sec:1}Introduction}

Ferroelectrics (FEs) have plethora of applications from actuators to sensors and from ultrasonic generators to energy-harvesting devices due to its piezoelectricity (or electric-field-induced strain) \cite{Scott2007,uchino2000,park1997,priya2009}. The piezoelectric coefficients $d$, which quantify the piezoelectric strain against an applied electric field, are the ubiquitous figures of merit in such applications \cite{park1997, saito2004, uchino2000}. Optimization of these parameters in single crystals and nanoparticles compounds with the existing deleterious effects such as twinning and depoling though many of them exhibit enhancement of piezoelectricity in certain nonpolar directions \cite{park1997, wada2005,kuwata1982}. (Here nonpolar in the sense that a direction other than the spontaneous polarization direction of the crystal). Nonetheless, FEs in the polycrystalline form are preferred over the single crystals in engineering owing to the ease in manufacturing and the compositional modifications of polycrystals \cite{Damjanovic1998}. Noncollinear
polarization rotation has been proposed as the possible origin of high piezoelectric response in FE single crystals \cite {fu2000,park1997,noheda2001, wu2003}.  Furthermore, phenomenological thermodynamic theory addressing the piezoelectric anisotropy relates it to the flattening of the free energy function in certain nonpolar directions \cite {budimir2005,budimir2006}.

As-grown polycrystalline FE is an aggregate of single crystalline grains with randomly oriented (spontaneous) polarizations \cite {lines1977}. The spatial configuration of crystallographic grains and their orientation distribution (texture)  impact the piezoelectricity  exhibited by conventional as well as new generation FEs \cite{Scott2007}. The randomness in polarization-vector orientation renders the resultant piezoelectricity of the material to be marginal or zero. Although the resultant polarization is zero for as-grown polycrystal, an overall piezoelectricity can be enabled by the application of an external electric field, called poling field, though all the grains can never align perfectly \cite {lines1977}.

The aggregate texture  of an unpoled polycrystal would have a uniform random distribution of orientation \cite{lines1977,brosnan2006}. With the strength of the poling field increases, we assume the nature of the grain orientation distribution to become Gaussian (normal) \cite {uetsuji2004,ruglowsky2006}. Piezoforce microscopic image studies of Pb(Zr,Ti)O$_3$ (or PZT) films leads also to a quantitatively similar kind of distribution of domains \cite {Awu2005}.   Spherical harmonics based orientation distribution function (ODF) is also
 used to model the texture of polycrystalline ferroelectrics \cite {rajapakse2007,Li2000,
 Lu1999}.  The texture analysis of  PZT by synchrotron X-ray diffraction reveals that the diffraction peak intensity ratio $R_{\{200\}}$ shows a strong dependance on $\sin^2\theta$, where $\theta$ represents the orientation angle between the plane normal and the polar axis of the material \cite {hall2004}.
Moreover, the domain distribution has been treated statistically to be in an orientation distribution characterised with March-Dollase function in X-ray and neutron diffraction studies of piezoelectric materials \cite{rogan2003}.
In this work, Rogan et al. \cite{rogan2003} has done  \textit{in situ} neutron diffraction study of the polycrystalline PZT while mechanical loading. This work provides  insights into the ferroelastic switching of domains in FEs and uses a March-Dollase function to model the evolved texture 
during loading. 

Electrical and mechanical loading (or poling) have limitations
 in realising smooth saturation textures due to the restrictions imposed
by the crystallographic symmetry \cite {jljones2005, Kamlareview2001}. Many domains cannot be
reoriented due to a complex set of internal stresses and electric fields in grains and because some domains will switch back after the poling field is removed \cite {Damjanovic1998}. Yet, the Gaussian distribution is found to realize the piezoelectricity especially at the boundaries bounded by a random distribution (where the piezoelectric coefficients, $d_{j\nu}\rightarrow 0$) and a perfect single crystal texture (where $d_{j\nu}$ tends to the single
crystal values) correctly \cite {jaya2009c}. This can overcome, for instance, the inadequacies of other distribution functions in approximating the case of strong preferred orientation, which eventually reflect in the correct evaluation of piezoelectricity \cite {vitalij2005}. For
instance, March-Dollase function models the distribution very efficiently
at low and moderate degrees of non-randomness. But the strong preferred orientation
cannot be adequately approximated by this function \cite {vitalij2005}. In the present model we present configuration of crystallites with orientation distribution fits into a Gaussian. Despite this, we are not deterministic about the methods (e.g., hot forging, reorientation by poling, tape casting) to realise  this grain configuration (distribution).  Nor are we attempting to model the evolution texture of piezoelectric ceramics in this paper albeit predicting
the configuration of grains that would optimize the piezoelectricity. 

The spatial and orientational randomness  of grains can be judiciously employed in the design of FE polycrystals (ceramics) with tailor-made configurations \cite {jayachandran2008}. Recently it is shown that overall piezoelectricity would be enhanced by the introduction of either [110]- or [111]-oriented grains into a random BaTiO$_{3}$ polycrystal \cite{Wada2007,Takahashi2007}. The piezoelectric coefficient $d_{33}$ scales up until a certain grain size and further decrease from this point would decrease the $d_{33}$ in these experiments. Though the ceramics possess randomly oriented grains, fine domains must be introduced to achieve this narrow window of high piezoelectric activity.  The solution of the optimal grain configuration of the polycrystalline FE is hardly resolved although much research activity is underway in ferroelectrics. This is mainly due to the vast number of possible configurations available at hand albeit the FE ceramics are easy to manufacture. In this work we would identify an optimum configuration of grains in the microstructure of a polycrystalline FE material with an enhanced piezoelectricity from both a textured ceramic and a single crystal. Moreover, we would discuss on the mechanism of polarization rotation in FEs in view of their piezoelectric response in the context of the optimization results. 

 For many FE single crystals, the polarization $\bm{P}$ rotates with an electric field applied along a nonpolar direction  and the polarization rotation in such cases does not necessarily evolve through a monoclinic phase \cite {fu2000}. The polarization  $P_{i}=d_{i\mu}T_{\mu}$, where T is the stress and $d$ piezoelectric strain coefficient. Here the Latin indices range from 1 to 3 and Greek indices from 1 to 6, owing to the Voigt contraction of tensorial
indices \cite {nye1985}. The typical value of the piezoelectric coefficient $d_{33}$ measured for  tetragonal (T) BaTiO$_{3}$ is $\approx$ 90 pC/N when [001]-poled while $d_{33}\approx$ 203 pC/N for [111]-poled engineered-domain single crystal \cite {zgonic1994, wada1999}. The piezoelectric coefficient $d_{31}$ typically falls between  -33.4 and -62.0 pC/N for [001]- and [111]-poled single domain BaTiO$_{3}$ crystals \cite {zgonic1994, wada1999}. Since templated grain growth \cite {messing2004} enables the fabrication of textured ceramics with a fraction of oriented material, it is important to find out the degree of orientation as well as the fraction of aligned grains in a matrix of random polycrystalline FE. Pb-free (K$_{0.5}$Na$_{0.5}$)(Nb$_{0.97}$Sb$_{0.03}$)O$_{0.3}$ polycrystal grown using this method is recently shown to possess excellent electromechanical properties if it is  $\langle001\rangle$-oriented with a narrow orientation distribution \cite {yunfei2009}. Such a possibility to fabricate useful polycrystals is one of the motivation to a continuum model study, as is shown in this paper.      

The orientation of a piezoelectric crystal is modeled by a set of Euler
angles $(\phi, \theta, \psi)$. Euler angles are defined in the following ways: first, the crystal is rotated by angle $\phi$ around the z-axis, then rotate an angle $\theta$ around the new x-axis, and finally by an angle $\psi$ around the new z-axis. All the rotations are in the counterclockwise
direction\cite {goldstein78}. The matrix of transformation from the 
crystallographic
coordinate system to a local coordinate system $\bm y$ is given by \cite
{goldstein78} 
\begin{widetext}
\begin{equation}
a_{ij}=%
\begin {pmatrix}
\cos\psi \cos\phi-\cos\theta \sin\phi \sin\psi&
\cos\psi \sin\phi + \cos\theta \cos\phi \sin\psi&\sin\psi \sin\theta\\
-\sin\psi \cos\phi-\cos\theta \sin\phi \cos\psi&-\sin\psi \sin\phi + 
\cos\theta \cos\phi \cos\psi&\cos\psi \sin\theta \\
\sin\theta \sin\phi&-\sin\theta \cos\phi&\cos\theta
\end{pmatrix}
\label{eq:1}
\end{equation}
\end{widetext}
The crystal orientation and thus the piezoelectric properties would therefore
inextricably depends on three Euler angles. Even in the case of single crystals
it would be tedious to survey  all the crystallographic orientations and
map the corresponding piezoelectric properties. This situation demands an optimization procedure to realize a particular objective function in a piezoelectric material. [There would be $73^{3}$ (= 389,017) different combinations of ($\phi,\theta,\psi$) if one chose angles at an interval of $5^\circ$
between $\pm \pi$ in FE single crystals alone]. Nonetheless, the case of polycrystals is too complicated as it is constituted ideally by thousands of single crystalline grains. To arrive at an optimum texture of the ferroelectric polycrystal at which the material exhibits maximum piezoelectric performance, a global optimization method has to be employed here as well. This is because the piezoelectricity depends on the parameters which controls the orientation distribution of the grains. Nevertheless, the choice of the optimal set of parameters is complicated and it is impossible to analyze all possible combinations of the distribution parameters or the angles themselves. We chose a modified stochastic global
optimization technique incorporating a generalised Monte Carlo scheme for
this purpose. A modified simulated annealing (SA) is quite suitable in this respect as the objective function is not sensitive to the starting point of the iterative process (the so called connectivity, where any state of
the system can be reached starting from any other state) \cite {vechi1983}.

We propose a design strategy based on a continuum mechanics and modeling to attain electromechanic figures of merit in FEs combined with a modified simulated annealing for optimization, in this paper. The model is developed
for crystals of all classes. In this paper, we apply this methodology to optimize the piezoelectricity in the classic perovskite, the tetragonal 
$P4mm$ BaTiO$_3$. The piezoelectric performance is quantified through the evaluation of the effective electromechanical properties of the FE single crystals and polycrystals. We have used the mathematical homogenization method \cite {guedes1990,silva1999} which efficiently characterizes the equilibrium macroscopic electromechanical properties of a polycrystalline ferroelectric material. 

\section{\label{sec:2}Model}
\subsection{\label{sec:2a}Homogenization of ferroelectrics}
The electrical and elastic boundary conditions and the orientation of the polarization and permittivity axes of grains (in polycrystals) and
crystals complicate the microscopic analysis of the FEs \cite{Landis2002,Landis2003}. Considering these, the mathematical modelling (if we describe the rapidly varying material properties with equally rapidly varying functions) and the numerical analysis of these materials will become difficult and sometimes even intractable. To put it simply, homogenization of partial differential equations (PDEs) has as its main purpose to approximate PDEs that have rapidly varying coefficients with equivalent "homogenized" PDEs that (for instance) more easily lend themselves to numerical treatment in a computer. The homogenization method accommodates the interaction of different phases in characterizing both the macro- and the micro-mechanical behaviors. (For instance, in a polycrystalline material the crystallite manifests as a phase in this sense).  In homogenization theory the material is locally formed by the spatial repetition of very small microstructures
(unit-cells), when compared with the overall macroscopic dimensions. Further, the material properties are periodic functions of the microscopic variable, where the period is very small compared with the macroscopic variable. This enables the computation of equivalent material properties by a limiting process wherein the microscopic cell size is approaching zero. 

The key idea in micromechanical modeling
is to relate the effective properties of a material to the properties
of its constituent parts, which may be the phases of a composite or
the grains of a polycrystal in ferroelectrics \cite {Huber2005,Landis2004,
HwangMcmeeking1999, Huber1999,Kamlareview2001, Li2000}.
A  review on the various methods for obtaining effective properties of ferroelectrics is given in an earlier paper \cite {jaya2009b}. The asymptotic analysis and homogenization of the piezoelectric medium has resulted in the equilibrium
macroscopic piezoelectric properties in tensor notation as follows;
\begin{eqnarray}
e_{prs}^H(\bm{x})=\frac{1}{|Y|}\int_{Y}\bigg[e_{kij}(\bm{x},\bm {y})\Big(
\delta_{kp}+\frac{\partial R^p}{\partial y_k}\Big)\Big(\delta_{ir}\delta_{js}+ \nonumber \\
\frac{\partial \chi_{i}^{rs}}{\partial y_j}\Big)-e_{kij}(\bm{x},\bm {y})
\frac{\partial \Phi _{i}^p}{\partial y_j} \frac{\partial \Psi^{rs}}{\partial y_k} \bigg] dY
\label{eq:2} \\
\kappa_{pq}^{\varepsilon H}(\bm{x})=\frac{1}{|Y|}\int_{Y}\bigg [\kappa_{ij}
^{\varepsilon}(\bm{x},\bm{y})\Big (\delta_{ip} + \frac {\partial R^p}{\partial
y_i}\Big )\Big (\delta_{jq} + \frac {\partial R^q}{\partial y_j}\Big ) 
\nonumber \\
-e_{kij}(\bm{x},\bm{y})\Big (\delta _{kp}+\frac {\partial R^p}{\partial
y_k}\Big )\frac {\partial \Phi^p_i}{\partial y_i}\bigg]dY
\label {eq:3} \\
C_{rspq}^{EH}(\bm {x})=\frac{1}{|Y|}\int_{Y}\bigg [C_{ijkl}^{E}(\bm{x},\bm
{y})\Big (\delta_{ip}\delta_{jq}+\frac{\partial \chi_{i}^{pq}}{\partial y_j}
\Big )\nonumber \\
\times \Big(\delta_{kr}\delta_{ls}+\frac{\partial \chi_{k}^{rs}}{\partial y_l} \Big ) \nonumber \\
+ \ e_{kij} (\bm{x},\bm{y})\Big (\delta_{ip}\delta_{jq}+\frac
{\partial \chi_{i}^{pq}}{\partial y_j} \Big )\frac
{\partial \psi^{rs}}{\partial y_k}\bigg ]dY.
\label {eq:4}
\end{eqnarray}
Here the $e$, $\kappa$ and $C$ are the electromechanical coefficients, viz.,
piezoelectric, dielectric and elastic stiffness coefficients respectively. Also, $\chi_{i}^{rs}(\bm{x},\bm{y})$ is the characteristic displacement, $R(\bm{x},\bm{y})$ is the characteristic electric potential, $ \Phi_{i}^{p}
(\bm{x},\bm{y})$ and $\psi_{rs}(\bm{x},\bm{y})$ are characteristic coupled functions of the FE unit-cell of size $Y$, satisfying a set of microscopic equations \cite {silva1999,jaya2007}. $\delta$ is the Kronecker delta symbol. (We have assumed Einstein summation convention that repeated indices are implicitly summed over throughout this paper). Symmetry requires that $e_{kij} = e_{kji}$, $\kappa_{ij}^{\varepsilon}= 
\kappa_{ji}^{\varepsilon}$ and $C_{\mu\nu}^{E}=C_{\nu\mu}^{E}$. 
In the above Eqs.~(\ref{eq:2})-(\ref{eq:4}) $e_{kij}(\bm{x},
\bm{y})$, $\kappa_{ij}^{\varepsilon}(\bm{x},\bm{y})$ and $C_{\mu\nu}^{E}
(\bm{x},\bm{y})$ are the electromechanical properties of the single crystallite whose collection constitutes the microscopic unit-cell and can be described in microscopic coordinates $\bm y$ as 
 
\begin {eqnarray}
\begin{array} {c c c}
C_{ijkl}^E&=&a_{ip}a_{jq}a_{kr}a_{ls}C_{pqrs}^{E{\prime}} \\
\kappa_{ij}&=&a_{ip}a_{jq}\kappa_{pq}^{\prime} \\
e_{ijk}&=&a_{ip}a_{jq}a_{kr}e_{pqr}^{\prime} 
\end {array}
\label {eq:5}
\end {eqnarray}
where $a_{ij}$ are the Euler transformation tensors from crystallographic
coordinate system to the local microscopic coordinates $\bm y$ \cite 
{goldstein78}. Here the primed moduli are the ones expressed in crystallographic coordinate system. Also the piezoelectric response is determined along an arbitrary crystallographic direction determined by the Euler angle 
$(\phi,\theta,\psi)$ with respect to the reference frame of the microstructure, i.e., $\bm y$. [Here the microstructure refers to the representative
volume element (RVE) of the ferroelectric polycrystal used for the homogenization]. The (resultant) polarizations of the crystallites in an as-grown polycrystal are randomly oriented in the lattice space and hence require three angles to describe its orientation with reference to a fixed coordinate system. Euler angles $\phi$, $\theta$ and $\psi$  can completely specify the orientation of the crystallographic coordinate system embedded in crystallites and thereby the orientation of   relative to a fixed Cartesian coordinate system. (The superscript $H$ would be dropped from the homogenized piezoelectric properties given in Eqs.~(\ref{eq:2})-(\ref{eq:4}) for brevity from
the rest of the discussion). 

After asymptotic analysis a set of microscopic system of equations characterising $\chi_{i}^{rs}(\bm{x},\bm{y})$, $R(\bm{x},\bm{y})$, 
$ \Phi_{i}^{p}(\bm{x},\bm{y})$ and $\psi_{rs}(\bm{x},\bm{y})$ is obtained
and is solved computationally. The three-dimensional (3D) numerical model developed is implemented in finite element method (FEM). The polycrystal sampling is performed using a unit-cell of $14\times14\times14$ mesh with 21,952 Gaussian integration points.  The convergence of piezoelectric properties with unit-cell size allows us to determine the simulation-space-independent, macroscopic piezoelectric properties at various distribution of grains. Also the convergence studies are necessary as the orientations (Euler angles) of the crystallites are chosen randomly from a Gaussian distribution of angles. A normal random generator delivers different sets of numbers each time it is invoked and this will affect the consistency of the results. Nonetheless, at a $14^3$ mesh we found that the statistical fluctuations on $d_{i\mu}$ are less than $4\%$ (Ref.~\onlinecite {jaya2009c}). After using the usual approximations of FEM, the set of linear equations for each load case is obtained where each global stiffness,
piezoelectric and dielectric matrix is the assembly of each element's individual matrix, and the global force and charge vectors are the assembly of individual force vectors for all the elements.

Full integration (2-point Gaussian integration rule in each direction) is used for the evaluation of the stiffness, piezoelectric and dielectric matrices and for the homogenization. As the representative microstructure (unit-cell) is expected to capture the response of the entire piezoelectric system, particular care is taken to ensure that the deformation across the boundaries of the cell is compatible with the deformation of adjacent cells. Thus all the load cases are solved by enforcing periodic boundary conditions in the unit-cell for the displacements and electrical potentials. Though the model is general
and is designed to accommodate all crystalline symmetries, we apply it to the case of BaTiO$_{3}$. The numerical homogenization of ceramic BaTiO$_{3}$ is carried out using the single crystal data taken from Ref.~[\onlinecite{zgonic1994}]
using the present homogenization model.

\subsection{\label{sec:2b} Optimization of piezoelectricity}

The texture of a polycrystalline ceramic can be quantified through the distribution
parameters corresponding to the angle of orientation in space. Euler angles $(\phi, \theta,\psi)$ can completely specify the orientation of the crystallographic coordinate system embedded in crystallites and thereby their orientations
with respect to a local coordinate system $y_i$. Since the aggregate texture for polycrystalline FEs follows a Gaussian distribution, the probability distribution function (pdf) is defined by,
\begin{equation}
f(\alpha\mid\mu,\sigma)=\frac{1}{(\sigma\sqrt{2\pi})}
\exp-\Big[\frac{(\alpha-\mu)^2}{2\sigma^2}\Big]
\label {eq:6}
\end{equation}
about the direction of the electric field. $\mu$ and $\sigma$ are the parameters of the distribution viz., the mean and the standard deviation respectively. $\alpha$ stands for the Euler angles $(\phi, \theta,\psi)$. Since $\mu$ and
$\sigma$ decide the scatter of orientations (of the grains) and for that matter be critical to the piezoelectric response of an FE ceramic, they would assume the role of design variables of the optimization problem. Thus we are aiming to find an optimum set of these parameters from a solution space controlled by the laws of coordinate transformations from a crystallographic coordinate system embedded in the grains to a local coordinate system which coincides with the global frame of reference. Also, the solution space is bounded by distribution parameters ranging from those of uniform (in the case of random polycrystal) to those of normal distribution (in the case of poled polycrystal). A fairly uniform distribution can be achieved by putting standard deviation ($\sigma$) equals 5 and for a poled ceramic ferroelectric the $\sigma$ is set near zero.

Here the objective is to search possible ways of enhancing the piezoelectricity in FE polycrystals. When an electric field is applied in a piezoelectric crystal the shape of the crystal changes slightly. This is known as the converse
piezoelectricity. There exists a linear relationship between the components of the electric field vector $E_i$ and the components of the strain tensor $\varepsilon_{ij}$ which describe the change in shape \cite {nye1985}. The
piezoelectric strain constant $d_{i\mu}$, (which is written in tensor form
as $\varepsilon_{jk}=d_{ijk}E_{i}$) thus quantifies the piezoelectricity
of a FE material. In a ceramic FE each of the three Euler angles $\phi$, $\theta$ and $\psi$ if observed individually falls in respective normal distributions thanks to the misalignment of the crystallographic axes of the constituent grains. Hence there would be six (3 pairs) parameters altogether viz., 
$(\mu_{\phi},\sigma_{\phi})$, $(\mu_{\theta},\sigma_{\theta})$ and 
$(\mu_{\psi},\sigma_{\psi})$ quantifying the scatter of angles from $y_i$. The grain distribution parameters chosen by the optimization algorithm will prompt a normal random generator and thereby create a set of Euler angles ($\phi,\theta, \psi$) for each of the grains. These Euler angles will dictate the coordinate transformation [as given by Eq.~(\ref {eq:5})] in the electromechanical property tensors appearing in the homogenization equations Eqs.~(\ref{eq:2}) - (\ref{eq:4}).

Each iteration of the optimization algorithm calls the objective function,
the effective piezoelectric coefficient $d_{j\mu}$.  The numerical solution of the coupled piezoelectric problem sought using the FEM would be substituted in the homogenization expressions. Nonetheless, the electromechanical tensor
undergo coordinate transformation dictated by the Euler angles as depicted
in the expressions in Eqs.~(\ref{eq:5}) before it enter the homogenization
algorithm. The FEM used for this study correlates each randomly oriented grain in a polycrystalline material with each element of the finite element mesh. The macroscopic piezoelectric coefficients $e_{i\mu}$ obtained in the homogenization equation Eq.~(\ref{eq:2}) is calculated first and subsequently
we calculate $d_{i\mu}=\sum_{\mu=1}^6e_{i\nu}s_{\mu\nu}^E$, by making use of the elastic compliance $s_{\mu \nu}^E$ derived from inverting homogenized $C_{\mu \nu}^E$ from Eq.~(\ref{eq:4}).   

The key idea behind simulated annealing is based in the Monte Carlo step proposed in the Metropolis algorithm \cite {Metropolis1953} for simulating the behavior of an ensemble of atoms that are cooled slowly from their melted state to the minimum energy ground state. The ground state or minimum energy state corresponds to the global optimum we are seeking in material optimization. In order to apply this algorithm to a piezoelectric material, we must first introduce the notion of \emph {system energy}. In the present setup the piezoelectric coefficient $d_{33}$ acts as system energy and we seek its  maximization. In order to be consistent with our definition of design variables, let 
\begin{eqnarray}
E(R_{i})&\equiv& d_{33}(\alpha) \Rightarrow \nonumber \\
E(R_{i})&\equiv& d_{33}(\sigma_{\phi},\mu_{\phi},
\sigma_{\theta},\mu_{\theta},\sigma_{\psi},\mu_{\psi})
\label {eq:7}
\end{eqnarray}
be the surrogate of energy $E$ of a particular configuration $R_{i}$. Here
each set of distribution parameters are selected randomly from $\sigma\in [0, 5]$ and $\mu\in [0,\pi/2]$. The main goal of optimization is to find the ground state(s), i.e., the minimum energy configuration(s), with a relatively small amount of computation. Minimum energy states are those that have a high likelihood of existence at low temperature. The likelihood that a configuration, $R_{i}$, is allowed to exist is equal to the Boltzmann probability factor, 
$P(R_{i})=exp[-\frac{E(R_{i})}{k_{B}T}]$, where $k_{B}$ is the Boltzmann
constant and $T$ is the temperature. $k_{B}$ is often treated as unity for
computational convenience \cite {weck2004}. 

\begin{figure}
\includegraphics [scale = 0.3]{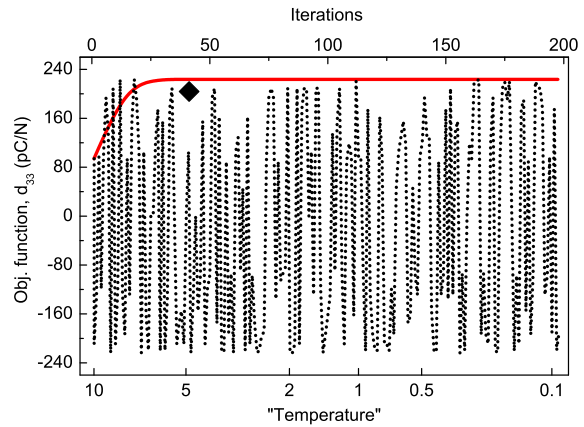}
\caption{\label{fig:1} (a) Objective function (piezoelectric coefficient $d_{33}$) as a function of \emph {temperature} in FE single crystal BaTiO$_3$. The filled diamond indicates the experimental value (= 203 pC/N) obtained in Ref.~\onlinecite {wada1999} for [111]-oriented engineered-domain single
crystal BaTiO$_3$. The dotted line shows the evolution of $d_{33}$ with iteration.
(Here
the \emph{temperature} is a decreasing control parameter of the optimization).}
\end{figure}

A control parameter similar to the temperature in physical annealing is introduced in optimization which will dictate the number of states to be accessed in going through the successive steps of the optimization algorithm before being settled in the minimum energy state (the optimum configuration). In single crystals, the design variables are the three Euler angles ($\phi,\theta,\psi$)
itself. The design domain is constrained by $-\pi\leq(\phi,\theta,\psi)\leq\pi$.
This condition would be enough to scan the entire crystallographic space
to search for the enhanced piezoelectricity. 

\begin{figure}
\includegraphics [scale = 0.3]{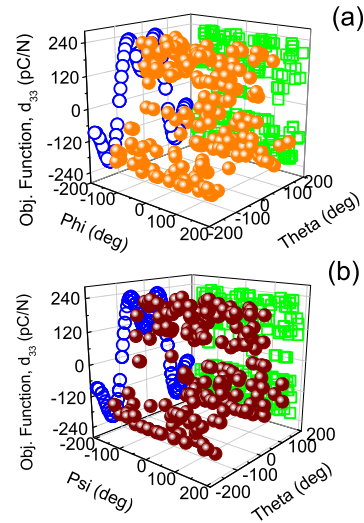}
\caption{\label{fig:2} The piezoelectric coefficient $d_{33}$ (spheres)  derived in the optimization as a function of Euler angles; (a) $\theta$ and $\phi$ and (b) $\theta$ and $\psi$.}
\end{figure}

The optimization problem can be summarised as to find $(\gamma)$ that maximize,
\begin {equation}
\left.
\begin{array}{lr}
\quad f(\gamma)\equiv d_{33} \\
\text{subject to,} \\
~ :-\pi \leq \gamma\equiv(\phi,\theta,\psi) \leq \pi   
\qquad ~ \text{for single crystal}  \\
\\
\left.
\begin{array}{lr}
: 0 \leq \gamma\equiv(\mu_\phi,\mu_\theta,\mu_\psi) \leq \pi/2 \\  
: 0 \leq \gamma\equiv(\sigma_\phi,\sigma_\theta,\sigma_\psi) \leq 5 
\end{array}
\right \}
\quad \text{for polycrystal}
\end{array}
\right \}
\label{eq:8}
\end {equation}

An interface between our modified simulated annealing algorithm (implemented in Matlab) and homogenization program (implemented in Fortran) is created to call $d_{33}$ (the objective function) at each iteration of the optimization. The temperature $T$ is set to fall geometrically by 20{\%} from each of the previous step $k$, i.e., $T_{k+1} = 0.8T_{k}$. Ideally, we must start the iteration with an initial guess of the design variables randomly picked up from $\sigma \in[0,5]$ and $\mu\in[0,\pi/2]$. To verify the correctness of the algorithm, the optimization procedure is first applied to the case of single crystal BaTiO$_{3}$. Thus we started with discrete Euler angles $\phi$, $\theta$, and $\psi$ alone without going to the assumption of distribution of grain orientations since a single crystal is devoid of any grain structure \cite {lines1977}. All the three angles are allowed values between limits $-\pi\leq(\phi,\theta,\psi)\leq+\pi$. This is to accommodate the analysis of orientational dependence of piezoelectricity in single crystal BaTiO$_{3}$ as well. 
\begin{table}
\caption{\label{tab:table1}Values of the homogenized piezoelectric strain
coefficients $d_{j\nu}$(in pC/N), piezoelectric stress
coefficients  $e_{j\nu}$ (in C/m$^2$) and free dielectric permittivity  $\kappa_{ij}^T$ (in $\kappa_0$)
of single crystal BaTiO$_3$. ($\kappa_0$ is the permittivity of free space).}
\begin{ruledtabular}
\begin{tabular}{lcccccccr}
 Method &$d_{15}$& $d_{31}$ &  $d_{33}$ &$e_{15}$& $e_{31}$ &  $e_{33}$& $\kappa_{11}^T$ & $\kappa_{33}^T$\\
\hline \\
Simulation & 560.7 & -33.7 & 94.0 & 34.2 & -0.7&6.7 & 4366 & 132\\
Experiment\footnote {Ref. \onlinecite {zgonic1994}}& 564.0 & -33.4 & 90.0 & 34.2 &-0.7&6.7& 4380&129
\end{tabular}
\end{ruledtabular}
\end{table}

\section {\label{sec:3} results and discussion}
\subsection {\label{sec:3a} Objective: single crystal piezoelectricity}

\begin{figure}
\includegraphics [scale  = 0.40]{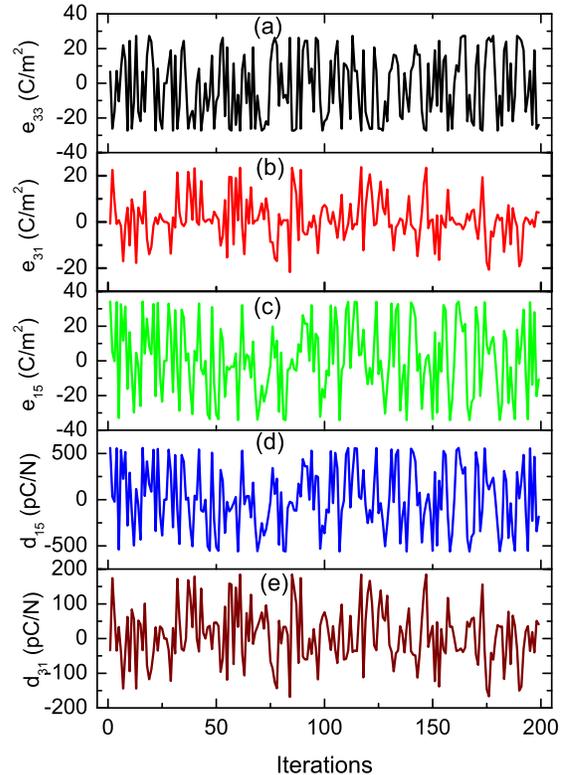}
\caption{\label{fig:3} Variation of effective piezoelectric stress coefficients (a) $e_{33}$, (b) $e_{31}$ and (c) $e_{15}$ along with piezoelectric strain
coefficients (d) $d_{15}$ and (e) $d_{31}$ in single crystal BaTiO$_3$ against iteration number.}
\end{figure}

Since the $d_{33}$ component (in tensor form $d_{33}$ is expressed as 
$\varepsilon_{33}=d_{333}E_3$, where $\varepsilon_{33}$ is the piezoelectric strain along the electric field $E_3$) of piezoelectric coefficient tensor better describes the piezoelectricity along the spontaneous
polarization direction of the BaTiO$_{3}$ \cite {park1997,saito2004}, we designate it as the objective function for our first optimization problem.
First, we would check the robustness of the homogenization implementation
before starting with the optimization. Good agreement between our homogenization  results on single crystalline BaTiO$_3$ and the experiment by Zgonik et. al. \cite {zgonic1994} is as shown in Table ~\ref {tab:table1}.  The comparison on Table ~\ref {tab:table1} would provide  only a check to the correctness of the algorithm. In fact, in our simulation we use the elastic stiffness $C_{\mu\nu}$, piezoelectric stress constants $e_{i\nu}$, and clamped dielectric permittivity  $\kappa_{ij}^{\varepsilon}$   from the measurements of Zgonik et al.~\cite{zgonic1994}. The present simulation is limited in its detail in the sense that we have not accounted the domain structure of the grains since they are assumed to be composed of as single domain.
 Nonetheless, the extrinsic contributions associated with the displacement of domain walls at external fields has a profound influence on the dielectric,
mechanical and piezoelectric properties of ferroelectric materials \cite {Damjanovic1998, wada2005, bassiri2007,Mckinstry2006}. Moreover the size
effect of grains and domains are critical to the piezoelectricity of ferroelectrics
as well \cite {newnham1998,takahashi2006}. 

The evolution of the objective function $d_{33}$ with the \emph {temperature} in single crystal BaTiO$_3$ is shown in Fig.~\ref{fig:1}. The piezoelectric coefficient $d_{33}$ obtained after optimization ($d_{33}= 223.7$ pC/N) compares well with experimental
\cite {wada1999} (also see, Table ~\ref{tab:table2}) and other theoretical results \cite {damjanovic2006, liu2006}. The solution $\phi$, $\theta$ and $\psi$ are -2.182, 0.873 and -0.175 radians respectively. This if expressed in degrees  ($\phi=-125^{\circ}$, $\theta= 50^{\circ}$ and $\psi=-10^{\circ}$) would correspond to one of the $<111>$ directions of the BaTiO$_3$ single crystal along which the maximum piezoelectric coefficient of $d_{33} = 203$ pC/N is measured \cite {wada1999}. 

The plots on Figs.~\ref{fig:2}(a) and (b) reveals that the piezoelectric coefficient $d_{33}$ always confined within $\pm 223.7$ pC/N.  Here  $YZ$-projection (circles) on Figs.~\ref{fig:2} (a) and (b) shows peaks at 
$\theta=\pm 50^{\circ}$ and at $\pm 140^{\circ}$.   That $d_{33}$ here follows the same pattern shown in earlier studies \cite {du1999,budimir2006b} about the rotation ($\theta$) of the single crystal  provides an additional proof of the efficiency of the algorithm and show how comprehensive is the solution space. Also, it
can be seen that the [$XZ$-projection (rectangles) on Figs.~\ref{fig:2} (a)
and (b)]  angles $\phi$ and $\psi$ have no visible influence on the piezoelectricity of BaTiO$_3$. The $d_{33}$ exhibits both highs and lows irrespective of the values of $\phi$ and $\psi$ as dictated by the symmetric reduction of the transformation equation in
Eq.~(\ref {eq:1}). Also, since the spontaneous polarization in tetragonal
BaTiO$_3$ is along the [001] direction, any rotation of that axis of the
crystal affects the piezoelectricity of the crystal measured in a local frame
of reference. This is evident considering the fact that the Euler angle $\theta$
measures the amount of rotation of the [001] axis of BaTiO$_3$. (This aspect would be discussed in detail in the coming section). In other words, for single crystalline BaTiO$_3$ to display enhanced piezoelectricity, one should cut the single crystal at an angle
away from the polar axis.

\begin{table}
\caption{\label{tab:table2} Optimized piezoelectric coefficient
$d_{33}$ along with experimental values of  single crystal and polycrystal
BaTiO$_3$. }
\begin{ruledtabular}
\begin{tabular}{lcr}
&\multicolumn{2}{c}{Optimum objective function, $d_{33}$ (pC/N)}\\
Authors & Single crystal&Polycrystal \\
\hline \\
Present & 223.7 \footnotemark[1] & 270.7 \footnotemark[2] \\
Wada et al. \footnotemark[3] & 203.0&-  \\ 
Zgonik et al. \footnotemark[4] &90.0&- \\
Bechmann \footnotemark[5] &-& 191.0 \\
\end{tabular}
\end{ruledtabular}
\footnotetext[1]{Optimization is  achieved when the crystal reaches the
orientation Euler angles ($\phi= -2.182,\theta = 0.873,\psi = -0.175)$. This
is a $\{111\}$ orientation.}
\footnotetext[2]{Optimal microstructure is characterised by the orientation
distribution parameters ($\mu_{\theta}=0.785$, $\sigma_{\theta}=0.1$, 
$\mu_{\phi}=1.134$, $\sigma_{\phi}=3.3$, $\mu_{\psi}=0.873$,  $\sigma_{\psi}=4)$
}
\footnotetext[3]{Experiment in Ref.~\onlinecite {wada1999} for [111]-poled
single crystal BaTiO$_3$}
\footnotetext[4]{Experiment in Ref.~\onlinecite {zgonic1994} for [001]-poled single crystal BaTiO$_3$}
\footnotetext[5]{Experiment in Ref.~\onlinecite {bechmann1956} for poled polycrystalline BaTiO$_3$}
\end{table}

 The objective function $d_{33}$ attains the optimum value of 223.7 pC/N at a \emph {temperature} of 6.4 [Fig.~\ref {fig:1}] and at the thirteenth iteration. (We have listed the $d_{33}$ parameter obtained at the optimization in Table ~\ref{tab:table2} along with the experimental values). After the symmetric reduction of the coordinate transformation equation  $d_{kij}=a_{kp}a_{im}a_{jn}d_{pmn}$, where $a_{ij}$ are the elements of Euler transformation matrix in Eq.~(\ref {eq:1}), yields
\begin{equation}
d_{33}=(d_{15}^{\prime}+d_{31}^{\prime})\sin^{2}\theta \cos\theta+d_{33}^{\prime}\cos^{3}\theta
\label{eq:9}
\end{equation}
for tetragonal $4 mm$ symmetry of BaTiO$_3$. Here the primed coefficients
corresponds to that of the spontaneously polarized BaTiO$_3$. Almost 89 {\%} of the contribution to the enhanced $d_{33}$ is from the first term of this equation containing the shear constant $d_{15}$. i.e., while evaluating the terms in the above Eq.~\ref{eq:9}, we obtain the contribution
from the first term as 198.8 pC/N and the rest of 223.7 pC/N is derived from the last term $d_{33}^{\prime}\cos^{3}\theta$. We have used the simulation results of $d_{j\nu}^{\prime}$ from Table~\ref{tab:table1} for this calculation. This corroborates the notion of the relation between polarization rotation, shear constant $d_{15}$ and the eventual piezoelectric enhancement \cite {davis2007}. 

The tetragonal phase of BaTiO$_3$ has \cite{jayachandran2008}
\begin{equation}
d_{33}=2e_{31}s_{13}+e_{33}s_{33}
\label{eq:10}
\end{equation}
Table~\ref{tab:table3} provides the set of piezoelectric coefficients 
$d_{j\nu}$, $e_{j\nu}$  and compliance $s_{\mu\nu}$ obtained at the point where the $d_{33}$ attains the optimal value in BaTiO$_3$ single crystals. Obviously, the major contribution to the $d_{33}$ comes from the second product of the
above equation (Eq.~\ref{eq:10}). As one can see from Tables~\ref{tab:table1}
and \ref{tab:table3}, the piezoelectric coefficient $e_{33}$ jumps from 6.7
to 27.3 C/m$^2$. This big increase contributes the major part of the enhancement
of $d_{33}$. The variation of $e_{j\nu}$ and  $d_{j\nu}$ are shown in Fig.~\ref{fig:3}.
Fig.~\ref{fig:3} facilitates the understanding of the geometrical relations
contributing to the enhancements of piezoelectricity in BaTiO$_3$.
In addition to this as shown in Eq.~\ref{eq:9}, $d_{33}$ becomes a linear
combination of $d_{j\nu}$ after coordinate transformation. This aspect would also
contributes to the enhancement of $d_{33}$.

\begin{table*}
\caption{\label{tab:table3} Piezoelectric coefficients $d_{j\nu}$ (in PC/N), $e_{j\nu}$ (in C/m$^2$) and compliance $s_{\mu\nu}$ (in 10$^{-12}$ m$^2$/N) obtained at the point of optimal $d_{33}$ in BaTiO$_3$.}
\begin{ruledtabular}
\begin{tabular}{lcccccccccr}
Phase & $e_{33}$ & $e_{31}$&$e_{15}$&$d_{31}$&$d_{15}$&$s_{11}$&
$s_{12}$&$s_{13}$&$s_{33}$&$s_{44}$ \\
\hline \\
Single crystal&27.3 & -17.7 & 4.7 & -144.2 & 115.4 & 6.7 & -2.5 & -1.4 & 5.9& 17.9\\

Polycrystal&30.0 & -13.8 & 14.4 & -124.0 & 254.5 & 6.3 & -1.9 & -2.2 & 6.9
& 18.0\\
\end{tabular}
\end{ruledtabular}
\end{table*}

\begin{figure}
\includegraphics [scale  = 0.3]{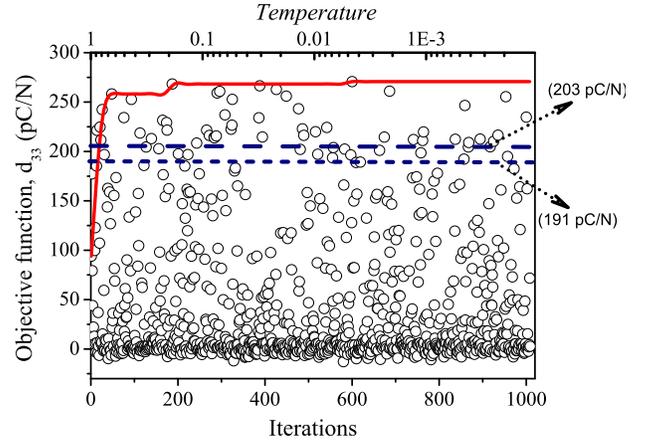}
\caption{\label{fig:4} Objective function (piezoelectric coefficient $d_{33}$) at each iteration in FE polycrystalline BaTiO$_3$. The circles indicate the value of $d_{33}$ at each iteration. The line shows the values retained by the optimization algorithm at each \emph {temperature} step. The dotted lines indicates the experimental values (= 203 pC/N) obtained for [111]-oriented engineered-domain BaTiO$_3$ (Ref.~\onlinecite {wada1999}) and (= 191 pC/N) for poled polycrystalline BaTiO$_3$ (Ref.~\onlinecite {bechmann1956}). (Here
the \emph{temperature} is a decreasing control parameter of the optimization). }
\end{figure}

\subsection {\label{sec:3b} Objective: polycrystal piezoelectricity} 

\begin{figure}
\includegraphics [scale  = 0.3]{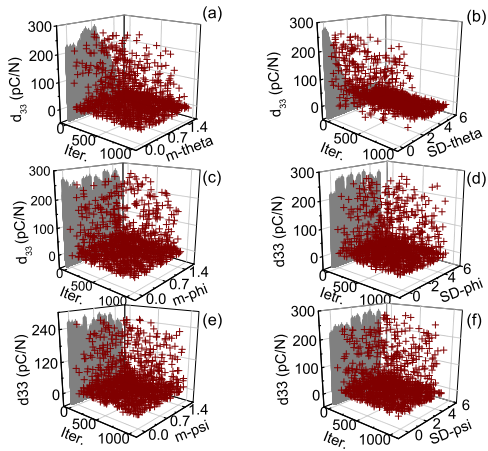}
\caption{\label{fig:5} Variation of the objective function, the piezoelectric
coefficient $d_{33}$ with the design variables, mean and standard deviation
of each of the angle distribution in polycrystal BaTiO$_3$. Each of the subplots
shows the $d_{33}$ against  the design variables, (a) $\mu_\theta$, (b)
$\sigma_\theta$, (c) $\mu_\phi$, (d) $\sigma_\phi$, (e) $\mu_\psi$ and   (f)
$\sigma_\psi$ corresponding to each iteration. }
\end{figure}

The optimization of polycrystal BaTiO$_3$ is treated in this section. The novelty in the present work is that the optimization procedure being applied to the ferroelectrics. Precisely, a stochastic optimization is developed used to find out the best grain configuration which will output a piezoelectric polycrystal wherein the piezoelectricity is maximum compared to the other possible configurations. We analyze the most general case with $(\sigma_{\phi},\sigma_{\theta},\sigma_{\psi})\in[0,5]$  and  $(\mu_{\phi},\mu_{\theta},\mu_{\psi})\in[0, \pi/2]$ as given by Eq.~\ref{eq:8}. The results are shown in Fig.~\ref{fig:4}. The initial and final temperature of the optimization is set to be 1 and 1.0634$\times$10$^{-4}$ respectively. Each step (there would be 42 steps to fall into the final temperature) constitutes 24 iterations and as a whole the final solution is realized in 1008 iterations. The objective function converges with a value $d_{33} = 270.7$ pC/N which is much higher than both [001] poled and [111] poled single crystal values
(see Table~\ref{tab:table2}). Here, the $d_{33}$ is obviously
enhanced by a factor of 3 from the [001] poled single crystal value of $d_{33} \approx90$ pC/N as is seen in Fig.~\ref{fig:4}. (Here the $d_{33}$ of textured polycrystal is being compared to that of the [001] poled single crystal). The optimal value of $d_{33}$ is also higher than the corresponding
value of an optimally oriented single crystal (where the value was $d_{33}$ = 223.7pC/N as shown in Table~\ref{tab:table2}). Also this is much higher than the poled polycrystal
experimental value of $d_{33} = 191$ pC/N obtained by Bechmann \cite {bechmann1956}. 

The solution   obtained is $\mu_{\theta}=0.785$, $\sigma_{\theta}=0.1$, 
$\mu_{\phi}=1.134$, $\sigma_{\phi}=3.3$, $\mu_{\psi}=0.873$,  $\sigma_{\psi}=4$.  It means $\theta$s (which measures the canting of the c-axis of the crystallographic grain) is kept at a small standard deviation of 0.1 but around a mean value of 0.785 radians (${\approx 45^\circ}$). Nevertheless, the other two angles are distributed with larger standard deviations close to the random.  Thus the present solution suggests one should keep the Euler angles $\phi$ and $\psi$ related to the orientation of ab-plane of the crystallites to be in random while the orientation $\theta$, of c-axes is kept close to ${45^\circ}$ but with a marginal standard deviation. A similar kind of result in polycrystalline BaTiO$_3$ was obtained numerically by Garcia et al.~\cite {garcia2005b}. They have shown that maximum piezoelectric response ($d_{33}$ or $d_{31}$) is exhibited by polycrystal ferroelectrics possessing specific crystallographic textures. 

\begin{figure}
\includegraphics [scale  = 0.45]{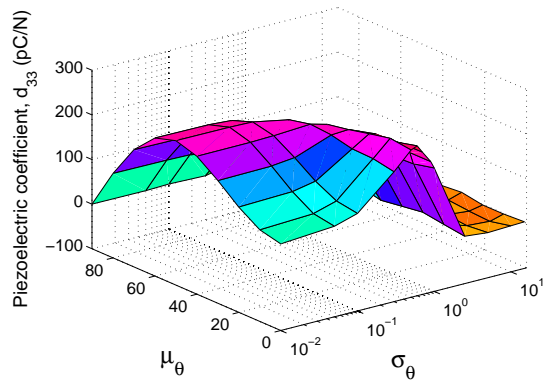}
\caption{\label{fig:6} Variation of the homogenized piezoelectric coefficient $d_{33}$  with the orientation distribution parameters $\sigma_{\theta}$ and $\mu_{\theta}$ (in degrees)  for the polycrystalline BaTiO$_3$. The other two Euler angles $(\phi, \psi)$  are kept at zero..}
\end{figure}

The dependance of the spread and shape of the orientation distribution on
the piezoelectric properties are   displayed in the three-dimensional plots
on Figs.~\ref{fig:5}. It can be seen that the effective piezoelectric constant
$d_{33}$ clearly shows a distinctive dependence on the distribution parameters
pertaining to the Euler angle $\theta$ [as can be seen from the pattern shown
in Figs.~\ref{fig:5} (a) and (b)]. (The contour on the $YZ$- projection on the plots shows the values of $d_{33}$ for the parameters $\mu_\theta$,
$\sigma_\theta$, $\mu_\phi$, $\sigma_\phi$, $\mu_\psi$, $\sigma_\psi$). The
piezoelectric coefficient follows a specific pattern for both $\mu_\theta$
and $\sigma_\theta$ shown on subplots Figs.~\ref{fig:5} (a) and (b) respectively.
 $d_{33}$ shows peaks around the mean $\mu_\theta\approx 0.7-0.8$ radians which would  be around ${\approx 45^\circ}$. Also, $d_{33}$ steadily increases
its value as the standard deviation $\sigma_\theta \rightarrow 0$, which
is clearly a tendency towards an aligned and textured ceramic material.
 
Next, we will study the influence of randomness on the enhancement of piezoelectricity
in ceramic BaTiO$_3$.  An important aspect observed in the optimization  study is that in polycrystals, unlike in single crystals, the orientation of the ab-plane of the
crystallites do play a role in determining its piezoelectricity. Another simulation is done to verify this point as shown in Fig.~\ref{fig:6}. Here the simulation is done keeping the $\phi$ and $\psi$ at zero while letting the $\mu_{\theta}$ and $\sigma_{\theta}$ varying. In contrast to the results obtained for polycrystal without any constraint (see Figs.~\ref {fig:4} and ~\ref {fig:5}), this simulation results in $d_{33}=236$ pC/N which is close to that of [111]-oriented single crystal. This would further point out that c-axes of the crystallites should be constrained to have a specific orientation while the ab-plane need not be kept at a specific texture but at random. This condition will deliver a ceramic piezoelectric material possessing better piezoelectricity than any other phase (optimally oriented single crystal or ceramic). Hence  randomness in the orientation of grains, if utilized judiciously, could be useful for manufacturing piezoelectric ceramics which outperform single crystals. Another point to noted here is the role played  by the shear constant
$d_{15}$ in the enhancement of $d_{33}$. Here the analysis of the the shear constant in the enhancement of piezoelectric response is complicated because of the presence of grain boundaries in polycrystal FEs. Yet, it would be important to know the value of $d_{15}$ to shed light on the origin of piezoelectric enhancement.  The value of the shear constant obtained at the maximum of $d_{33}$ is $d_{15}$= 254.5 pC/N (see Table~\ref{tab:table3}). As in single crystals the enhancement in $d_{33}$ in polycrystals is accompanied by a sizeable $d_{15}$ value in BaTiO$_3$.  A similar conclusion
was drawn by Garcia et al.~\cite{garcia2005b}, where they show that in partially
textured ceramics with a $d_{15}$ larger than $d_{33}$ is a good choice for
enhancement of polycrystal $d_{33}$.

\begin{figure}
\includegraphics [scale  = 0.35]{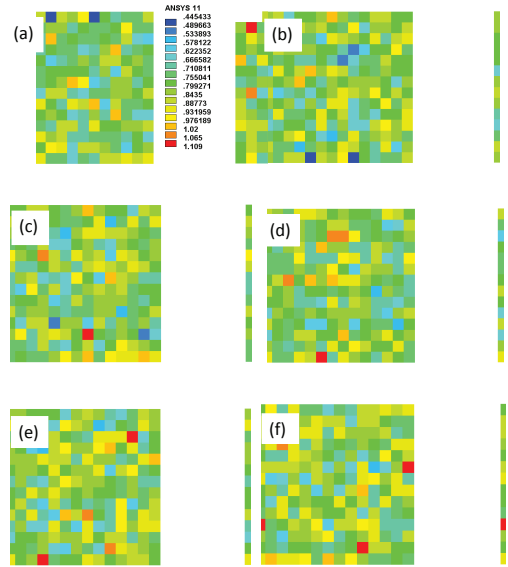}
\caption{\label{fig:7} The distribution of Euler angle $\theta$ in
optimized polycrystalline BaTiO$_3$. Here the angles in radians are mapped in colors. The six planes shown from (a)-(f) are (100), (001), $(\overline{1}00)$, (010), $(0\overline{1}0)$ and $(00\overline{1})$ respectively.}
\end{figure}

The optimal orientation distribution across the six faces of the microstructure (representative volume element) is shown in Fig.~\ref{fig:7}. Since the angle $\theta$ influences the $d_{33}$ the most, that only is shown. It is seen that (Fig.~\ref{fig:7}) the angles are mostly clustered around 0.8 radians (${\approx 45^\circ}$). The effective piezoelectric coefficient is dictated by the transformation of coordinates involving the Euler angles [see Eqs.~(\ref {eq:1}) and (\ref {eq:5})]  and hence is influenced by the rigorous combinations provided by the randomness. One of the possible reasons for higher piezoelectricity in FE ceramics with a certain pattern of grain distribution is the role played by the intrinsic polarization of the grains. The misalignment of the polarization in the neighboring grains would impart residual stresses, which couples with the electrical field to enhance the polarization and thereby the piezoelectricity. 
In the optimal distribution, the polarizations of the 
juxtaposed grains are configured in such a way that they will contribute
positively to the overall increase of piezoelectricity of the polycrystal.

In summary, we have introduced a global optimization technique to identify
the ideal configuration of both single and polycrystal FE. This method can
be used to the FEs of any symmetry and to any objective function involving
electromechanical coefficients. Here this method is applied to the perovskite,
tetragonal ferroelectric BaTiO$_3$. The paper predicts configuration of crystallites which maximizes the piezoelectricity of the polycrystalline ferroelectrics. The orientation distribution resulted from electrical and
mechanical poling is limited by many factors \cite{Kamlareview2001}. The alignment
of domains with the poling field is by switching its polarization to an equilibrium
position defined by the symmetry. The as-grown polycrystal before  poling has  domains randomly oriented. Thus the poling doesn't
yield the polarization vectors (of domains) all aligned along the field since  there may not exist an equilibrium orientation for some domains along the
 poling field direction. Apart from being provide a guide to experiment, essentially it doesn't suggest a particular method (for instance, mechanical or electrical poling, tape-casting, templated grain growth, hot
forging etc.) to realise the predicted crystallite configuration. A plausible approach is to adopt a combination of suitable processing method (to achieve a certain degree of preferred orientation in the unpoled state) and poling.  

In this work, we have observed significantly enhanced piezoelectric response in ferroelectric polycrystals at certain grain configurations. In single crystal BaTiO$_3$ the piezoelectricity is found to be larger along a nonpolar direction away from the polar axis.    We have optimized the ferroelectric ceramic by design at the microstructure level for piezoelectric applications. The solution obtained from the optimization procedure results in a three-fold enhancement of piezoelectricity in the ceramic phase compared to the single crystalline phase. If we use the randomness of the grain orientations judiciously the ceramic can replace even the oriented single crystals in piezoelectricity. A plausible reason behind the anisotropy
shown by both rotated single crystal and polycrystal FE could be the macroscopic
symmetry. The crystallographic symmetry is characterised by the anisotropic
(or isotropic) physical properties. The product phase obtained after the crystal orientation might be different from the parent phase in symmetry. The insight obtained from the optimization have the potential
to advance the design and discovery of complex FE configurations with superior piezoelectric performance. Further studies in this direction in relaxor ferroelectrics, where the single crystals display larger piezoelectricity,  could inaugurate new possibilities in technological applications which involves the requirement of high piezoelectricity.

\begin{acknowledgments}
KPJ acknowledges the award of Ci{\^e}ncia 2007 by FCT, Portugal. 
\end{acknowledgments}


\end{document}